%% file: manuscript.tex
\documentclass[final,5p,twocolumn]{elsarticle} 
\input{PackagesAndPreamble}

\begin{document}
\title{Early Career Developers' Perceptions of Code Understandability.\\A Study of Complexity Metrics}

\author[OULU]{Matteo Esposito}
\ead{matteo.esposito@oulu.fi}

\author[UNIBZ]{Andrea Janes}
\ead{andrea.janes@unibz.it}

\author[TAU]{Terhi Kilamo}
\ead{terhi.kilamo@tuni.fi}

\author[OULU]{Valentina Lenarduzzi}
\ead{valentina.lenarduzzi@oulu.fi}

\address[OULU]{University of Oulu, Finland}
\address[UNIBZ]{Free University of Bozen-Bolzano, Italy}
\address[TAU]{Tampere University, Finland}

\begin{abstract}
\textbf{Context.} Code understandability is fundamental. Developers need to understand the code they are modifying clearly. A low understandability can increase the amount of coding effort, and misinterpreting code impacts the entire development process. Ideally, developers should write clear and understandable code with the least effort.

\noindent\textbf{Aim.} Our work investigates whether the McCabe Cyclomatic Complexity or the Cognitive Complexity can be a good predictor for the developers' perceived code understandability to understand which of the two complexities can be used as criteria to evaluate if a piece of code is understandable. 

\noindent \textbf{Method.} We designed and conducted an empirical study among 216 early career developers with professional experience ranging from one to four years. We asked them to manually inspect and rate the understandability of 12 Java classes that exhibit different levels of Cyclomatic and Cognitive Complexity. 

\noindent\textbf{Results.} Our findings showed that while the old-fashioned McCabe Cyclomatic Complexity and the most recent Cognitive Complexity are modest predictors for code understandability when considering the complexity perceived by early-career developers, they are not for problem severity. 

\noindent\textbf{Conclusions.} Based on our results, early-career developers should not be left alone when performing code-reviewing tasks due to their scarce experience. Moreover, low complexity measures indicate good understandability, but having either CoC or CyC high makes understandability unpredictable. Nevertheless, there is no evidence that CyC or CoC are indicators of early-career perceived severity.Future research efforts will focus on expanding the population to experienced developers to confront whether seniority influences the predictive power of the chosen metrics. 
\end{abstract}


\maketitle

\input{Section/Introduction.tex}
\input{Section/RelatedWork.tex}

\input{Section/Background.tex}

\input{Section/EmpiricalStudy.tex}

\input{Section/Results.tex}

\input{Section/Discussion.tex}
\input{Section/ThreatsToValidity.tex}
\input{Section/Conclusion.tex}

\bibliographystyle{elsarticle-num-names}
\bibliography{manuscript}

\end{document}

%% file: PackagesAndPreamble.tex
\journal{Information and Software Technology}

\usepackage[utf8]{inputenc}
\usepackage[T1]{fontenc}
\usepackage{amsmath}
\usepackage{mathrsfs}
\usepackage{algorithmic}
\usepackage{graphicx}
\usepackage{textcomp}
\usepackage{footmisc}
\usepackage{tablefootnote}
\usepackage{multirow}
\usepackage{url}
\usepackage{caption} 
\usepackage[export]{adjustbox}
\usepackage{enumitem}
\usepackage{listings}
\usepackage{booktabs}
\usepackage{soul}
\usepackage{xcolor}
\usepackage{fourier} 
\usepackage[many]{tcolorbox}
\usepackage{fontawesome}

\usepackage{todonotes}
\newcommand\ra[1]{\textcolor{black}{#1}}

\AtBeginDocument{%
  }

\tcbset{
    sharp corners,
    colback = white,
    before skip = 0.2cm,    
    after skip = 0.5cm      
}                           
\newtcolorbox{boxC}{
    colback = sub, 
    boxrule = 0pt,  
}

\definecolor{main}{HTML}{CFCFCF}    
\definecolor{sub}{HTML}{CFCFCF}     

\newcounter{keyTakeAwaysCounter}

\newenvironment{keyTakeAways}[1][Key Take Away]
    {
    \addtocounter{keyTakeAwaysCounter}{1}
        \begin{boxC}
        \faLightbulbO ~ \thekeyTakeAwaysCounter. \textbf{#1}.\\
        }{
        
        \end{boxC}
}

%% file: Section/Introduction.tex
\section{Introduction}
\label{Intro}

Code understandability is an essential aspect of software development.
Code understandability can be defined as the measure to which ``code possesses the characteristic of understandability to the extent that its purpose is clear to the inspector''~\cite{Boehm1976}. Having poor code understandability in the program code can increase the amount of coding effort by more than 50\%~\cite{Minelli2015, Xia2018}.
To avoid misinterpretation of the code, developers should write code that requires the least amount of effort to be understood~\cite{Munoz2020}. 

When code is easy to understand, developers can more easily identify and fix errors, modify existing code, and integrate new code into existing projects. On the other hand, code that is difficult to understand can lead to confusion, errors, and time-consuming troubleshooting \cite{Tashima2018}.
Several factors contribute to code understandability, including the use of clear and concise syntax, consistent formatting, naming conventions, as well as well-or\-ga\-nized code structure. Additionally, documentation and comments can also play a crucial role in improving code understandability.
Different metrics, such as the McCabe Cyclomatic Complexity~\cite{McCabe1976} (CyC) and the CoC Complexity~\cite{Campbell2018} (CoC), have been proposed in the past to evaluate the complexity of the code. 
Current static analysis tools allow developers to keep track of these metrics in their code in real-time. 
CoC has been introduced by \ra{the tool} SonarQube\footnote{\label{Sonar} \url{https://www.sonarqube.org}} as an extension of the McCabe CyC, to evaluate code understandability~\cite{Campbell2018} better. 
The effect of CoC on code understandably was investigated by two recent studies by Scalabrino et al.\cite{Scalabrino2019} and Munoz~\cite{Munoz2020}. 
CoC seems to be a good indicator of understandability where a higher value means a reduction of understandability. 
Moreover, as highlighted by Munoz~\cite{Munoz2020}, the different complexity and understandability metrics are not deeply investigated and validated. In particular, it is still not evident which metrics better support the prediction of code understandability~\cite{Scalabrino2019}.
As a consequence, Lavazza et al.~\cite{Lavazza2022} extended the work of Munoz et al.~\cite{Munoz2020} \ra{comparing} CoC and CyC to identify which metric provides an advantage for code understandability. 

However, Code can also be complex due to problems such as design issues or code smells. As highlighted by Politowski et al.~\cite{Politowski2020}, anti-patterns in the code can decrease the code understandability and increase the effort needed to modify the code. Therefore, if the complexity metrics are correlated with code understandability, problems in the code can also be associated with the complexity measures.  

Previous studies highlighted the need to understand whether CoC is correlated with understandability better than the other existing metrics, and the previous results, based on mining software repository studies, could not tip the scales. Therefore, we decided to investigate the impact of these two metrics on code understandably from the point of view of the developer's perception.

We designed and conducted an empirical study involving 216 early-career developers with at least one year of experience. We asked them to manually inspect twelve Java classes to provide their opinion on the understandability of the code, on the presence of issues, and to rate the severity of the existing problem, if any. We analyzed the developers' answers and correlated the developer's perceived understandability with the CyC and CoC.

Moreover, suppose a positive but lower correlation exists between complexity measures and code understandability. In that case, we also aim to understand if complexity measures are correlated with the developers' perceived severity of problems in the Java code.

While there are differences between developers' opinions on the perception of the complexity of the code, the overall data indicate that CoC and CoC are modest indicators of the perceived understandability of the code. We also found that neither CyC nor CoC is correlated with the presence and the perceived severity of problems in the code. 

We \textbf{focused on early career developers} because we believe it is important to grasp their perceptions. For instance, (1) identifying how they interpret and use the complexity measures in real-world coding scenarios gives insight into where theoretical knowledge might not find its way into practice. Moreover, such insights can also show (2) how more intuitive and effective training programs, tools, and resources can be designed to suit them better. 

Therefore, our work can aid in highlighting common challenges and misbeliefs in the first years of a professional career, hence, professional growth, thus marking possible venues to select good practices that might improve the quality and maintainability of the code in general, boosting overall productivity and confidence for early-career developers. 

\textbf{Paper Structure}. Section~\ref{RW} describes the related work, in Section~\ref{Background} we introduce the background of this work, while in Section~\ref{EmpiricalStudy} we outline the research methodology adopted in this study. Section~\ref{Results} presents and discusses the obtained results. Section~\ref{Threats} identifies the threats to validity and Section~\ref{Conclusion} draws the conclusions.

%% file: Section/RelatedWork.tex
\section{Related Work}
\label{RW}
Code understandability is described as the measure of how well ``code possesses the characteristic of understandability to the extent that its purpose is clear to the inspector''~\cite{Boehm1976}. To better understand a piece of code, legibility is one of the main factors to take under control since if code is harder to read, it could be harder to understand~\cite{Boehm1976}.
Code understanding requires building high-level abstractions from code statements, visualizations, or models~\cite{Storey2000,Lin2006}. 
However, also readable code could be difficult to understand~\cite{Scalabrino2019}. 

Code understandability can be measured by considering several different factors. One possibility is based on the perceived understandability reported by developers answering comprehension questions~\cite{Dolado2003, Salvaneschi2014}, or filling out blank program parts \cite{Borstler2016}, or extending and/or modifying existing pieces of code~\cite{Hofmeister2017}.
To be more accurate, some studies traced the time to perform the assigned task, both questions or developing ones~\cite{Aljunid2012,Hofmeister2017}.
Other approaches evaluate code understandability focusing on physiological metrics detected by biometrics sensors~\cite{Fucci2019,Yeh2017} or eye-tracking devices~\cite{Fritz2014,Turner2014}. 
Moreover, considering the perceived understandability by rating the different pieces of code under analysis, can provide a positive step forward in this field~\cite{Scalabrino2019}. Different factors can positively or negatively influence how developers perceive the understandability of a piece of code~\cite{Scalabrino2019}, which can be useful to develop a model to automatically measure the understandability. 
Several studies investigated the role of software metrics focusing on complexity as well as source-level metrics, such as LOC~\cite{Scalabrino2019} and CyC~\cite{Kasto2013,Scalabrino2019} or CoC~\cite{Campbell2018} during the developing process or maintenance tasks~\cite{Feigenspan2011}. Moreover, other types of metrics such as documentation-related metrics such as comment readability and metrics relating to a developer's experience were considered by researchers~\cite{Scalabrino2019}.
Results showed that none of the investigated metrics accurately represent code understandability~\cite{Feigenspan2011,Scalabrino2019}. However, all the software metrics considered in these studies suffered from empirical validation of their ability to measure code understandability. In particular, CoC needs more accurate validation~\cite{Campbell2018}.
However, the results demonstrated that such metrics can improve the effectiveness of the code understandability under evaluation~\cite{Scalabrino2019}.

A deeper investigation of CoC has been performed by Munoz et al.~\cite{Munoz2020} and later by Lavazza et al.~\cite{Lavazza2022}. Munoz et al.~\cite{Munoz2020} considered CoC the metric measured by SonarQube and evaluated the association with different code understandability metrics: the time taken to understand a code snippet, the percentage of correctly answered comprehension questions on a code snippet, subjective ratings of a comprehension task, and physiological measures on the subjects engaged in understanding code.
Results showed that CoC is correlated with the time spent by a developer to understand source code. However, they did not compare the magnitude of this correlation against different complexity metrics. 
As Lavazza et al.~\cite{Lavazza2022} reported in their work ``\textit{before embracing the use of CoC, we need to understand whether CoC is correlated with understandability better than the measures that were proposed in the past for the same purpose}''. 
To assess it, Lavazza et al.~\cite{Lavazza2022} conducted an empirical study extending study \cite{Munoz2020}. They correlated CoC and CyC to identify which metric provides an advantage for code understandability. Unfortunately, the achieved results are not proposed for a particular metric.

%% file: Section/Background.tex
\section{Background}
\label{Background}

CyC was introduced by McCabe in 1976~\cite{McCabe1976}. It is a graph theoretical measure of program complexity. CyC  measures the amount of linearly independent paths in the program. It is based on the assumption that the more independent paths there are in a program, the more complex the program is likely to be.

The definition of CyC is based on representing program code as a control flow graph, i.e. a directed graph with all execution paths of the program depicted. Each node in the graph represents a basic code block and each edge is a pass of control between the blocks. Based on the graph, CyC $M$ is calculated as $M = E - N + P$ where $E$ is the number of edges, $N$ is the number of nodes, and $P$ is the number of strongly connected components in the graph. 
While CyC is a widely used metric to indicate the error proneness of program code, it fails to address certain code issues especially when it comes to computational complexity. CyC is poor at handling nested conditions and iterative structures \cite{Suleman2013}. It has been regarded as a poor metric for code understandability \cite{Campbell2018}.

In SonarQube, the Complexity measure is calculated based on the CyC of the code \cite{sonarcc} where each split in the control flow of a function increments the complexity measure by one. However, there are small differences between languages in how the complexity gets calculated due to differences in language structures.

CyC can be used as an indicator of how difficult a program is to test, maintain, or modify. Programs with high CyC are generally more difficult to understand, analyze, and change, as they contain more decision points and potential paths through the code. CyC is often used as a quality metric to evaluate the maintainability and overall complexity of software programs.

CoC is based on the idea that not all decision points in a program are equally difficult for a human to understand. Some decisions are simple and easy to reason about, while others are more complex and require more mental effort. CoC assigns a weight to each decision point in the code based on its level of complexity, with more complex decisions receiving a higher weight.

In SonarQube, CoC was introduced as ``a new metric for measuring the understandability of any given piece of code''~\cite{Campbell2018}. Based on the documentation \cite{sonarcogc}, CoC exhibits some similarity with CyC defined by McCabe~\cite{McCabe1976}, since CoC can address some of the ``common critiques and shortcomings belonging to CyC''~\cite{Suleman2013}. 
Moreover, CoC can fill the gap related to understandability present in the CyC~\cite{Crasso2014, Misra2018, Campbell2018}. 
Crasso et al.~\cite{Crasso2014}  suggested a set of metrics that combines various measurements encompassing essential aspects of object-oriented programming that have been validated by Misra et al.~\cite{Misra2018}. In both studies, the authors evaluated existing CoC metrics (such as Cognitive Functional Size and Cognitive Program Complexity Measure), chose those that highlight distinct features, made any required adjustments, and then assembled them into a unified collection.

SonarQube computed CoC metrics based on three basic rules~\cite{Campbell2018}: 
\begin{enumerate}
    \item ``Ignore structures that allow multiple statements to be readably shorthanded into one'';
    \item ``Increment for each break in the linear flow of the code'';
    \item ``Increment when flow-breaking structures are nested''.
\end{enumerate}






%% file: Section/EmpiricalStudy.tex
\section{Empirical Study Design}
\label{EmpiricalStudy}
This section presents the empirical study's goal, research questions, metrics, and hypotheses. We outline the study context, the data collection, and the analysis. We designed and conducted an empirical study following the guidelines proposed by Wohlin~\cite{Wohlin2000}. 

\subsection{Goal and Research Questions}
The \textit{goal} of this study is to compare CyC and CoC with the \emph{purpose} of understanding which complexity metric better represents the early career developer's perceived complexity of the Java code. 
The \emph{perspective} is of researchers since they are interested in understanding what complexity metrics can be more helpful to understand the code complexity. 

Based on our study goal, we derived the following Research Questions:

\begin{boxC}
\textbf{RQ$_1$.} What is the accuracy of early career developers in detecting problematic classes?
\end{boxC}

Understanding how early career developers perceive code understandability is important because early career developers are usually put to work on maintenance and evolution \cite{ahmad2024early,zhao2024early,esposito2024validate}. Hence, the ability of these developers to correctly identify classes with problems significantly impacts software quality and development efficiency \cite{meyer2019today,esposito2024correlation}. Therefore, we aim to evaluate the accuracy of those developers in their early years at detecting problem classes within software code.


Although important, assessing early-career development's accuracy would only lightly scratch the surface of the problem. No previous study has investigated possible proxies of developer's perception regarding perceived understandability and severity. Therefore, we computed complexity metrics such as CoC or CyC, and we asked:

\begin{boxC}
\textbf{RQ$_2$.} Can complexity metrics indicate an early career developer's perceived understandability of classes?
\end{boxC}
CyC and CoC are employed to measure the complexity. Hence the maintainability of software \cite{kafura1987use}. The unnoticed and un-foddered fact is how well such metrics go with the perception of understanding according to early-career developers, who have less experience and heuristics than their senior peers. Therefore, analyzing the relationship between the perceived understandability and complexity metrics could hint at whether current complexity metrics are useful predictors of how less-experienced developers view understandability.

In particular, since CoC is considered a ``more contextualized form of quantitative data on code complexity,'' we are interested to understand if CoC is a better predictor for code understandability. Since CoC was built upon the CyC, we hypothesized it might better represent the code understandability. Therefore we conjectured one \textbf{hypothesis} ($H_1$) as follows:

\begin{itemize}
    \item $H_{11}$: \textit{There is a correlation between complexity metrics and early career developer's perceived understandability of classes.}
\end{itemize}

\noindent Hence, we defined the \textbf{null hypothesis} ($H_0$) as follows:
\begin{itemize}
    \item $H_{01}$: \textit{There is no correlation between complexity metrics and early career developer's perceived understandability of classes.}
\end{itemize}

Complex code is considered hard to modify~\cite{Minelli2015,Xia2018}. Moreover, code affected by high levels of CyC is usually affected by more severe problems~\cite{Minelli2015, Xia2018}. Hence we ask:

\begin{boxC}
\textbf{RQ$_3$.} Can complexity metrics indicate an early career developer's perceived severity of an existing issue in a class?
\end{boxC}
As stated before, although complexity metrics are designed to quantify numerous facets of code that could impact its maintainability and readability, it remains an open question how the different complexity metrics align with the issue severity ranking of human developers in their early careers \cite{kafura1987use,ahmad2024early,esposito2023can}. This relationship can only be understood because early career developers usually identify and fix code issues in real projects \cite{ahmad2024early,esposito2023uncovering}. If complexity metrics could become a good predictor for perceived severity, they could be used as an orientation means, guiding these developers in their work. Hence, we aim to establish the strength of the relationship between the established complexity metrics and the severity of issues in code classes as perceived by early-career developers. Moreover, we considered that a lower code understandability could lead to a misleading in the problem identification in the inspected code and, consequently, a wrong perception of its severity. Therefore we conjectured one \textbf{hypothesis} ($H_2$) as follows:

\begin{itemize}
    \item $H_{12}$: \textit{There is a correlation between complexity metrics and early career developer’s perceived severity of an existing issue in a class.}
\end{itemize}

\noindent Hence, we defined the \textbf{null hypothesis} ($H_0$) as follows:
\begin{itemize}
    \item $H_{02}$: \textit{There is no correlation between complexity metrics and early career developer’s perceived severity of an existing issue in a class.}
\end{itemize} 

\subsection{Study Context}
As for participants, we selected junior developers. The reason for choosing them instead of senior developers is because they are the developers that most frequently need to approach new code. In particular, junior developers commonly need to extend existing code in their company, fix bugs, or integrate new features. Therefore, we selected early career developers from various Finnish software houses. 
We selected participants based on their current tasks, specifically working on existing code, and needed to understand problems in the code when extending it or fixing bugs. 
We finally involved 216 junior developers with Java experience ranging from one to four years. 

We did not present the CyC and CoC values to the participants to not influence their ability to recognize a potential design problem only because they see the complexity values in advance.

\subsection{Study Setup and Data Collection}

\begin{figure*}[]
     \centering
 \includegraphics[width=0.7\linewidth,trim={0 0 0 0},clip]{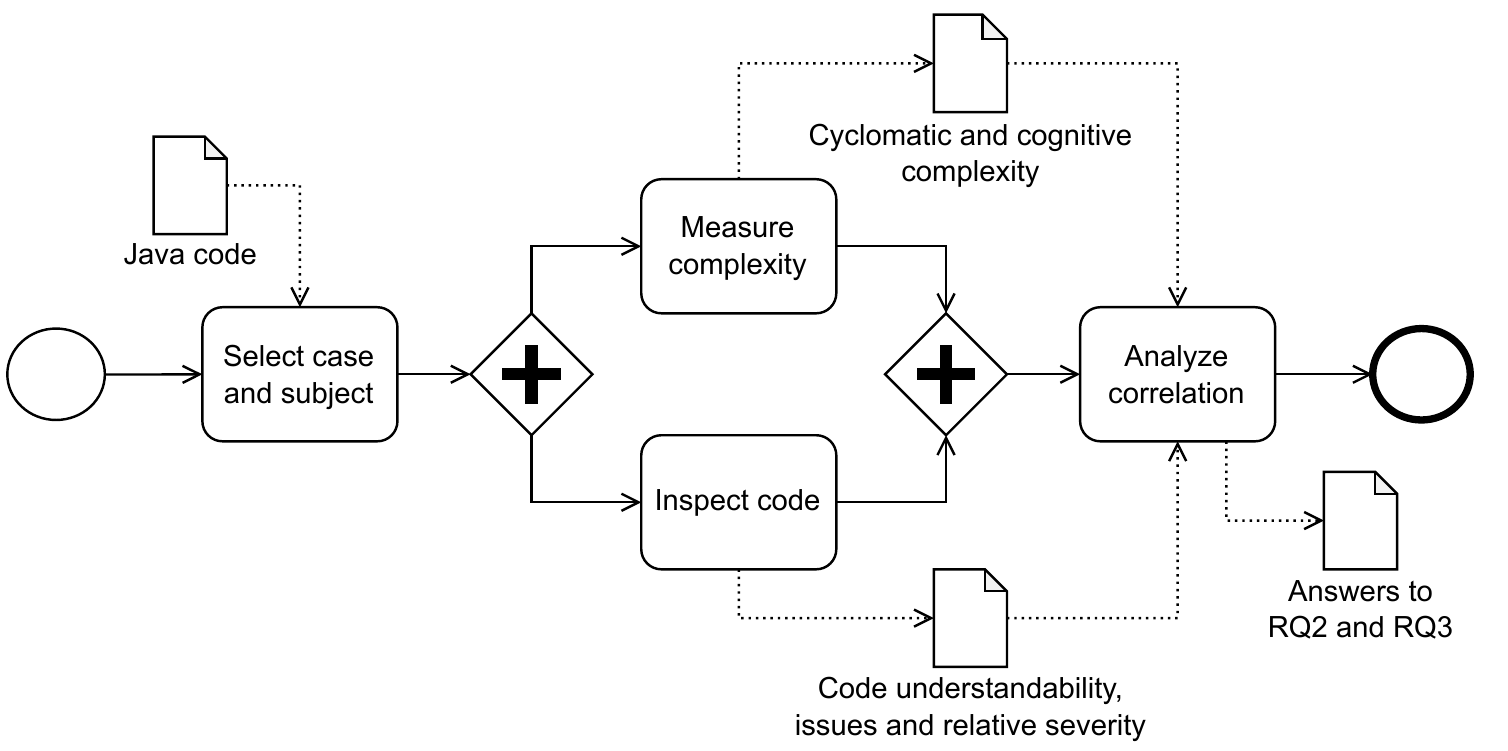}
     \caption{Empirical study design process} 
     \label{fig:Design}
 \end{figure*}

We designed our empirical study using the five steps below to answer our research questions. 
Figure~\ref{fig:Design} illustrates the process using the Business Process Model and Notation (BPMN) \cite{omg2011bpmn} specification language.

\begin{enumerate}
    \item \textit{Code Selection:} We selected Java code affected by problems of different severity from Apache Software Foundation projects. 
    \item \textit{Complexity measurement:} We measured the CyC  and CoC of the selected Java code using SonarQube. 
    \item \textit{Developers selection:} We identified the junior developers to be included in our study.
    \item \textit{Code inspection:} We asked developers to inspect the selected Java code and to provide their opinion on the understandability of the code, on the presence of issues, and to rate the severity of the existing problem, if any. 
    \item \textit{Data Analysis:} We analyzed the developers' answers and correlated the developer's perceived understandability with the CyC  and CoC.
\end{enumerate}

\subsubsection{Code Selection}
In this section, we report the case and subject selection for this study. 
We selected classes written in Java that were affected by different problems that can influence code understandably from Apache Software Foundation projects. Two of three authors and a senior Java developer independently evaluated the presence of issues in the code. Then, all three persons discussed possible inconsistencies and finally defined a list of 12 classes where all agreed on the presence of the same issues (Table~\ref{tab:Classes}).
More details of the selected classes and the problems identified in the code are available in the replication package\footref{package}. 
Table~\ref{tab:gt} details the selected classes and the problems identified in the code. All the authors inspected the selected classes and independently reported the identified problems. All the authors validated the identified problem at the end of the process.

\subsubsection{Complexity Measurement.} We measured the code complexity using CoC and CyC by applying SonarQube version 7.5\footref{Sonar}. 
We selected SonarQube since, to our knowledge, it is the only tool that detects both and reports how each metric is calculated. Moreover, we rely on SonarQube since it is one of the most adopted static analysis tools by practitioners \cite{vassallo2019developers, Avgeriou2020}.

\subsubsection{Code Inspection.} We asked developers to manually inspect the 12 Java classes and provide their opinions about the code understandability. 
To collect the information, we organized the questionnaire into four sections:

\begin{itemize}
\item \textit{Respondents' Background}. We collected the profiles of the respondents who were considering development experience.
\item \textit{Code Inspection}. In this questionnaire section, we asked participants to manually inspect a Java class and provide their opinion about their perceived \textit{Code Understandability} through a five-point Likert Scale (1 means ``very difficult'' and 5 means ``very easy''). The questions are reported in Table \ref{tab:Questions}. 
\item \textit{Perceived Problem Criticality} in the code, reporting if the problem exists in the class and rating their \textit{severity} through a five-point Likert Scale (1 means ``very low severity'' and 5 means ``very high severity'').
\end{itemize}

We implemented the questionnaire using Google Forms. The Questionnaire is available in the replication package\footref{package}.

\input{Tables/cyccoc}

\begin{table*}[]
    \centering
     \small
    \caption{Questions about code inspection's section}
    \label{tab:Questions}
    \begin{tabular}{l|p{10.5cm}|l}
\hline 
\textbf{RQ}& \textbf{Question} & \textbf{Answer type}\\ \hline 
RQ$_1$  & How easy is it to understand the code of this class? & Five-point  scales  \\  
\multirow{2}{*}{RQ$_2$} & Does, in your opinion, this class has design, coding style, or any other problems? &  Yes/No \\
& If YES, please rate the severity of the problem & Five-point Likert Scales \\ \hline 
\end{tabular}
\end{table*}


\subsubsection{Study Execution}
We provided the participants with instructions describing how to access the classes and answer the survey. The participants could inspect the classes and complete the online questionnaire in one round.
We informed the participants, according to the GDPR\footnote{https://gdpr-info.eu}, about their rights and that they could abandon the study anytime. Moreover, all information provided by each participant has been treated as confidential, without disclosing any sensible data.

\subsection{Data Analysis}
 This section describes the procedure for analyzing the data collected.

Concerning the results of the code inspection phase, we first verified the participants' backgrounds by analyzing the distribution of their education level (bachelor or master)  and their experience as developers in software companies.  

To address our research question, we quantitatively analyzed developers' reported perceptions of code understandability. 

To investigate developers' accuracy (RQ$_1$), we measured it with widely adopted IR accuracy metrics \cite{esposito2023uncovering,esposito2024extensive,esposito2024leveraging} (Table \ref{tab:metrics} and computed the agreement among developers, i.e., Inter-Rater Agreement (IRA) as suggested by established guidelines \cite{ralph2021empirical}.

\input{Tables/metrics}


We analyze developers' IRA via Fleiss's $\kappa$ \cite{fleiss1971measuring}. 
When assigning items to different categories, the kappa statistic is commonly used to evaluate the agreement between raters or classifiers. In our context, we used it to measure the level of IRA between the three HRs who reviewed the output of models and  HEs. When comparing the observed agreement level to the expected agreement level, Fleiss's $\kappa$ provides a metric for the reliability and consistency of the categorizations among two or more reviewers. We opted for Fleiss's $\kappa$ over Cohen's $\kappa$ because the latter is restricted to only two reviewers. Table   \ref{tab:kappa-agreement} presents the interpretation Fleiss's $\kappa$ \cite{fleiss1971measuring, 10.1093/ptj/85.3.257, esposito2024beyond}.

To investigate whether CyC and CoC (independent variables) can be indicators of a developer's perceived code understandability (RQ$_2$) and severity (RQ$_3$) (dependent variable), we assessed the distribution of the collected data to select the appropriate statistical tests for hypothesis testing and correlation analysis \cite{falessi2023enhancing,ccarka2022effort}. and  we conjectured one \textbf{hypothesis} ($H_\mathcal{N}$) as follows:

\begin{itemize}
    \item $H_{1\mathcal{N}}$: \textit{The data do not belong to a normal distribution.}
\end{itemize}

\noindent Hence, we defined the \textbf{null hypothesis} ($H_0$) as follows:
\begin{itemize}
    \item $H_{0\mathcal{N}}$: \textit{The data belongs to a normal distribution.}
\end{itemize} 

We tested $H_\mathcal{N}$ with the Anderson-Darling (AD) test \cite{10.1214/aoms/1177729437}. The AD test assesses whether data samples derive from a specific probability distribution, such as the normal distribution. AD measures the difference between the sample data and the expected values from the tested distribution. More specifically, it evaluates differences in the cumulative distribution function (CDF)  between the observed data and the hypothesized distribution \cite{10.1214/aoms/1177729437}.

The AD test and the Shapiro-Wilk (SW) test \cite{shaphiro1965analysis} are both statistical tests used to assess the normality of data. 

The SW test focuses on the correlation between the observed data and the expected values under a normal distribution, emphasizing the smallest and largest values in the dataset. Therefore, according to \citet{mishra2019descriptive}, SW is more appropriate when targeting small datasets ($\leq50$ samples). 
On the other hand, the AD test considers a broader range of values, including those in the middle of the distribution, providing a more sensitive evaluation of normality, especially for larger sample sizes than our own. 
Therefore, we preferred  AD over SW \cite{doi:10.1080/01621459.1974.10480196}. 
AD is also considered one of the most powerful statistical tools for detecting most deviations from normality \cite{doi:10.1080/01621459.1974.10480196,stephens2017tests}.

Test results allowed us to \textbf{reject the null hypothesis} for each test case, $H_{0\mathcal{N}}$, asserting that the \textbf{data was not normally distributed}. 

Therefore, we adopted the Spearman rank correlation coefficient $\rho$ instead of Pearson's to test the $H_{1}$ and $H_{2}$ values \cite{pearson1895notes}. Spearman's $\rho$ is a statistical measure used to assess the strength and direction of association between two variables, and it is a non-parametric measure that evaluates the monotonic relationship between variables. Therefore, $\rho$  assesses whether the variables tend to increase or decrease together, regardless of the exact rate of change. 


We interpret the values of $\rho$,according to Cohen~\cite{cohen1988statistical}: no correlation if $0 \leq \rho < 0.1$, small correlation if $ 0.1 \leq \rho < 0.3$, medium correlation if $0.3 \leq \rho < 0.5$, and large correlation if $0.5 \leq \rho \leq 1$. Corresponding limits apply for negative correlation coefficients.
We set our significance level to the standard value of $\alpha = 0.05$.

\input{Tables/Kappa}
\input{Tables/gt}

Specifically, regarding \textbf{RQ$_3$}, each author has conducted the qualitative data analysis individually.
Moreover, pairwise inter-rater reliability was measured across the three sets of decisions to get a fair/good agreement on the first iteration of this process.  Based on the disagreements, we clarified possible discrepancies and different classifications. A second iteration resulted in perfect agreement.  We conducted the analysis only for the cases where participants correctly identified a problem in the code.  

\subsection{Replicability}
\label{sec:Replicability}
To allow our study to be replicated, we have published the complete raw data in the replication package\footnote{\label{package}\url{https://doi.org/10.5281/zenodo.12743183}}.

%% file: Tables/cyccoc.tex
\begin{table}[]
    \centering
    \footnotesize
    \caption{Selected classes}
    \label{tab:Classes}
    \resizebox{\linewidth}{!}{%
\begin{tabular}{rrrrrrrrrrrrr}
\hline
\textbf{Class} &
  \multicolumn{1}{l}{\textbf{C1}} &
  \multicolumn{1}{l}{\textbf{C2}} &
  \multicolumn{1}{l}{\textbf{C3}} &
  \multicolumn{1}{l}{\textbf{C4}} &
  \multicolumn{1}{l}{\textbf{C5}} &
  \multicolumn{1}{l}{\textbf{C6}} &
  \multicolumn{1}{l}{\textbf{C7}} &
  \multicolumn{1}{l}{\textbf{C8}} &
  \multicolumn{1}{l}{\textbf{C9}} &
  \multicolumn{1}{l}{\textbf{C10}} &
  \multicolumn{1}{l}{\textbf{C11}} &
  \multicolumn{1}{l}{\textbf{C12}} \\ \hline
CyC &
  37 &
  130 &
  113 &
  56 &
  57 &
  35 &
  110 &
  23 &
  2 &
  6 &
  7 &
  3 \\
CoC &
  38 &
  73 &
  36 &
  68 &
  3 &
  6 &
  0 &
  6 &
  1 &
  0 &
  1 &
  0 \\ \hline
\end{tabular}%
}
\end{table}

%% file: Tables/metrics.tex
\begin{table}
  \centering
  \small
  \caption{IR Metrics}
  \label{tab:metrics}
 \begin{tabular}{l|c}
\hline
\textbf{Metric} & \textbf{Formula}                                                 \\ \hline
Precision       & $\dfrac{TP}{TP + FP}$                                            \\ & \\

Recall          & $\dfrac{TP}{TP + FN}$                                            \\ & \\
MCC              & $\frac{TP * TN - FP * FN}{\sqrt{(TP+FP)(TP+FN)(TN+FP)(TN+FN)}}$. \\ & \\
F1-score        & $\dfrac{2 \cdot Precision \cdot Recall}{Precision + Recall}$    \\ 
\\\hline 
\end{tabular}%
\end{table}

%% file: Tables/Kappa.tex
\begin{table}[tb]
\centering
\small
\caption{Interpretation of $\kappa$ values for measuring agreement}
\label{tab:kappa-agreement}
\begin{tabular}{ll}
\hline
Value of $\kappa$ & Interpretation \\
\hline
$\kappa < 0$ & No agreement \\

$0 \leq \kappa < 0.4$ & Poor agreement \\

$0.4 \leq \kappa < 0.6$ & Discrete agreement \\

$0.6 \leq \kappa < 0.8$ & Good agreement \\

$0.8 \leq \kappa < 1$ & Excellent agreement \\
\hline
\end{tabular}
\end{table}

%% file: Tables/gt.tex
\begin{table}
    \centering
    \small
    \caption{Class Ground Truth}
    \label{tab:gt}
    \begin{tabular}{ccp{0.5\linewidth}}
\hline
\text{\textbf{Class}} & \text{\textbf{Has Design Issue}} & \text{\textbf{Issue Description}} \\
\hline
\text{C1} & \text{Yes} & \text{-maintainability: code smells} \\
\hline
\multirow{2}{*}{\text{C2}} & \multirow{2}{*}{\text{Yes}} & \text{-CoC exceeded} \\
& & \text{-maintainability: code smells} \\
\hline
\text{C3} & \text{Yes} & \text{-untested} \\
\hline
\multirow{3}{*}{\text{C4}} & \multirow{3}{*}{\text{Yes}} & \text{-buggy code} \\
& & \text{-maintainability: code smells} \\
& & \text{-CoC exceeded} \\
\hline
\multirow{2}{*}{\text{C5}} & \multirow{2}{*}{\text{Yes}} & \text{-duplicated code} \\
& & \text{-maintainability: code smells } \\
&& \text{(constants missing)} \\
\hline
\text{C6} & \text{Yes} & \text{-duplicated code} \\
\hline
\text{C7} & \text{Yes} & \text{-maintainability: critical code smells} \\
& & \text{(unimplemented functions)} \\
\hline
\text{C8} & \text{Yes} & \text{-minor code smells} \\
\hline
\text{C9} & \text{Yes} & \text{-code smell: exception handling} \\
\hline
\text{C10} & \text{No} & \text{-untested, but fine} \\
\hline
\text{C11} & \text{No} & \text{-untested, but fine} \\
\hline
\text{C12} & \text{Yes} & \text{-minor code smell} \\
\hline
\end{tabular}
\end{table}

%% file: Section/Results.tex
\section{Results}
\label{Results}
In this section, we report the results obtained by answering our RQs. We collected information from 216 early-career developers.

\begin{table}
    \centering
    \small
    \caption{Background Information}
    \label{tab:BackgroundResults}
    \begin{tabular}{p{1cm}|p{1cm}|p{2.7cm}|p{1cm}}
    \hline 
    \multicolumn{2}{c|}{\textbf{Education level}} & \multicolumn{2}{c}{\textbf{Developer Experience}} \\ \hline
    Bachelor & 21\% & less than 2 years & 84\% \\  
    Master & 79\% &  3 and 4 years & 16\% \\ \hline
\end{tabular}
\end{table}

\subsection{Early-career developers accuracy and agreement}
In this subsection, we address background information on the early-career developers interviewed, their accuracy in classifying problematic classes, and their agreement.

\noindent\textbf{Background Information.} 
Table~\ref{tab:BackgroundResults} presents the respondents' education level and years of experience. More specifically, we interviewed 170 master's degrees (79\%) and 46 (21\%) bachelor degrees, of which 96 (79.34\%) had between 1 and 2 years of developing experience, while the remaining 25 (20.66\%) had more than 3 and 4 years (Table~\ref{tab:BackgroundResults}). 

\begin{figure}
    \centering
    \includegraphics[width=\linewidth]{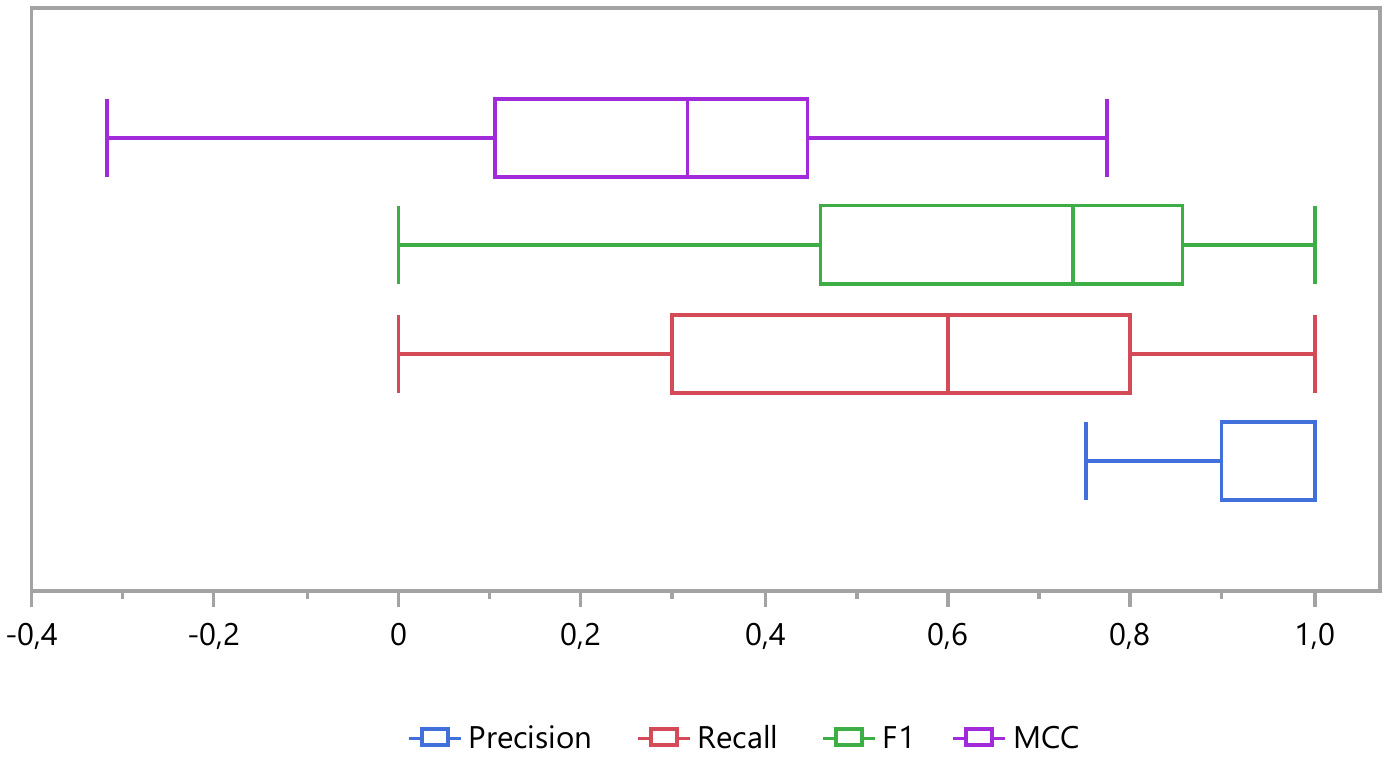}
    \caption{Distribution of early-career developers' accuracy in detecting problematic classes.}
    \label{fig:accuracy}
\end{figure}

\noindent\textbf{Accuracy.}
Figure \ref{fig:accuracy} presents the distribution of early-career developers' accuracy in detecting problematic classes in Precision, Recall, F1, and MCC. According to Figure  \ref{fig:accuracy}, we note that albeit developers have a promising Precision with a mean value circa 0.97, they fail to detect true positives, detecting a considerable number of false positives reflected in the Recall score that although scoring a mean value of 0.6, the interval ranges from 0 to 1. MCC exemplifies the imbalance between Precision and Recall with negative values as low as -0.35 with a mean value of 0.3.

\begin{figure*}
    \centering
    \includegraphics[width=\linewidth]{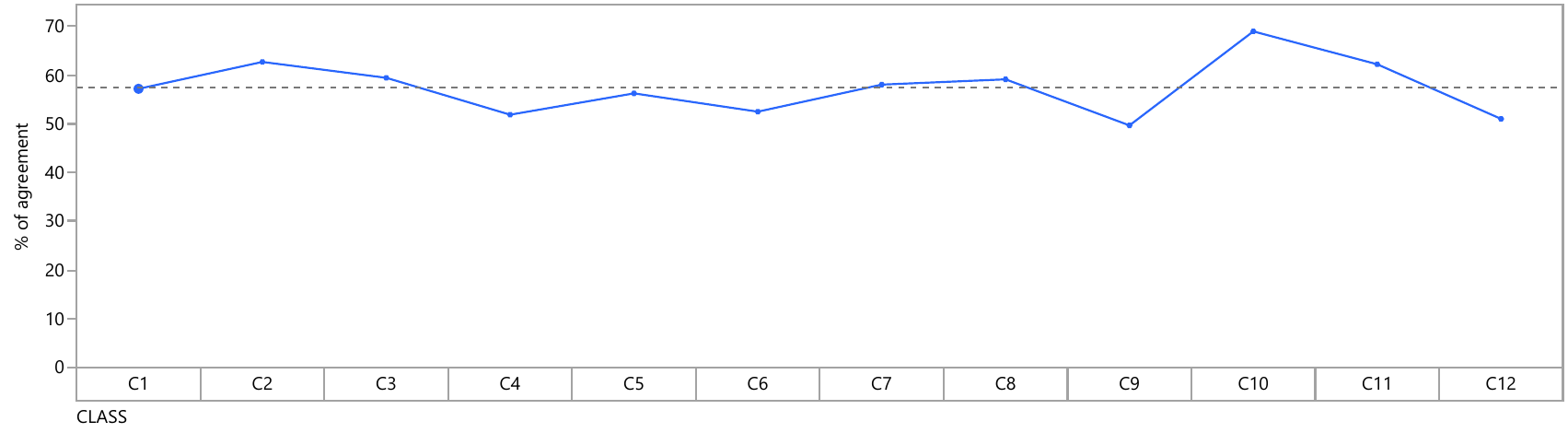}
    \caption{Average agreement percentage by Class.}
    \label{fig:IRA_CR}
\end{figure*}

\noindent\textbf{IRA.} Figure \ref{fig:IRA_CR} reports the average agreement percentage by Class and Reviewer. According to Figure \ref{fig:IRA_CR}, on average, the percentage of agreement per class is over  55\% with peaks of 67\% for C10 and as low as 49\% for C9. Conversely, the percentage of agreement per reviewer varies wildly between 45\% and 70\%.

\begin{figure}
    \centering
    \includegraphics[width=\linewidth]{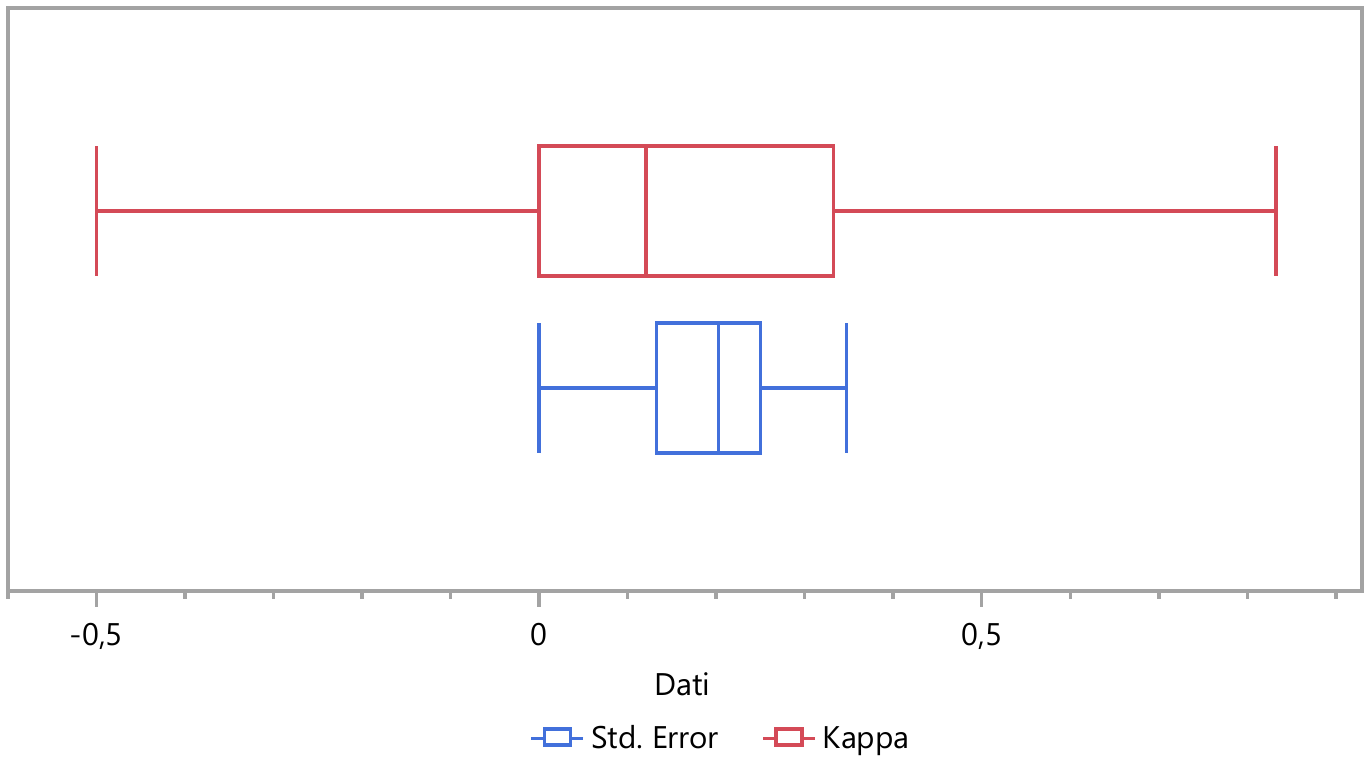}
    \caption{Fleiss's $\kappa$ and error standard deviation (RQ$_1$).}
    \label{fig:IRA}
\end{figure}

Therefore, to evaluate IRA, we used Fleiss's $\kappa$. Figure \ref{fig:IRA} reports Fleiss's $\kappa$ and the error standard deviation. According to Figure \ref{fig:IRA}, the IRA is dispersed between negative discreet agreement and good agreement with a positive poor agreement as mean value.
Hence, we can affirm that on average \textbf{early-career developers have different sensibility to issues in a class}.


Finally, to answer RQ$_1$, we can conclude that  \textbf{early-career developers have scarce agreement on classifying problematic classes with a discrete accuracy}.

\subsection{Complexity metrics and perceived code understandability (RQ$_1$)}

\begin{figure} [h]
    \centering
    \includegraphics[width=\linewidth]{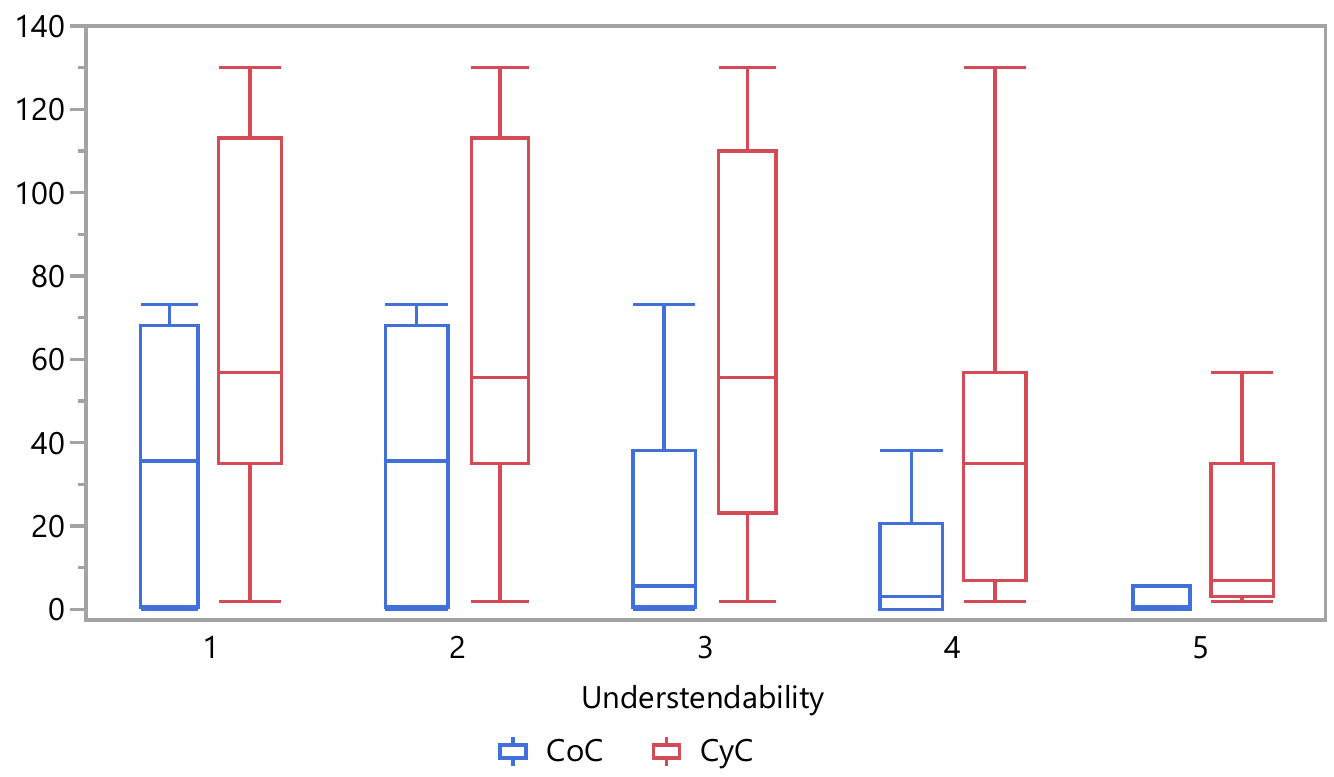}
    \caption{Distribution of complexity metrics values by the perceived code understandability (RQ$_2$)}
    \label{fig:Understandability}
\end{figure}

Figure \ref{fig:Understandability} presents the distribution of CyC and CoC by perceived understandability. We note the trend for which classes deemed Hard to understand correspond to lower values of CyC and CoC.

Table~\ref{tab:ResultsRQ1} presents the perceived code understandability. According to Table~\ref{tab:ResultsRQ1}
The respondents deemed half of the classes (C6-C12) very easy to understand, while the other half were neither hard nor easy.


Table~\ref{tab:spearman_understendability} presents Spearman's $\rho$ to test the correlation between CyC, CoC, and the perceived understandability. We note that the correlations are statistically significant for both complexity metrics. Therefore, we can affirm that  \textbf{CyC and CoC exhibit a negative, moderate correlation}. Hence, these results statistically confirm what is already presented in Figure \ref{fig:Understandability}, a \textbf{low level of understandability corresponds to higher values of the complexity metrics}.

\subsection{Complexity metrics and perceived severity of the design problem (RQ$_3$)}

\input{Tables/understendability}


\begin{table}
\centering
\small
\caption{Code Understandability - Correlation (RQ$_2$)}
\label{tab:spearman_understendability}
\begin{tabular}{l|r|r} \hline
\textbf{Spearman} &\textbf{CyC} & \textbf{CoC} \\\hline 
$\rho$ & -0,3819 & -0,3282 \\  
p-value & <0.0001 & <0.0001 \\ \hline 
\end{tabular}
\end{table}

\textbf{Perceived design or coding style problem.} The percentage of respondents who consider a class affected by a design or coding style problem is less than 75\% (Table~\ref{tab:ResultsRQ2RQ3}).

\textbf{Design coding style problem identification.} 
It is interesting to note that almost all the participants who identified a problem in the classes correctly identified at least one of the actual problems 
previously validated by all the authors. 
The only exception is for classes C7 and C12 not all the developers provided a description. 77.85\% of them correctly identified the correct problem for C7 and 89\% considering C12 class. 
Table~\ref{tab:ResultsRQ2RQ3} shows the results for identifying the problems grouped by Class (C).
Therefore, as also highlighted by Table~\ref{tab:ResultsRQ2RQ3}, 
we can conclude that the understandability of the code is independent of the perception of a problem.

\input{Tables/severity}

\begin{table}
\centering
\small
\caption{Perceived Problem Severity - Correlation (RQ$_3$)}
\label{tab:SpearmanSeverity}
\begin{tabular}{l|r|r} \hline
\textbf{Spearman} &\textbf{CyC} & \textbf{CoC} \\\hline 
$\rho$ & -0.2665 & -0.1686 \\  
p-value & <0.0001 & <0.0001  \\ \hline 
\end{tabular}
\end{table}

\textbf{Design or coding style problem severity.} The participants rated their concern for the design problem identified in the inspected code for each class. The participants rated their evaluation based on a 5-point Likert scale (1 means ``\textit{very low}'' and 5 means ``\textit{very high}''). 
Table~\ref{tab:ResultsRQ2RQ3} shows the obtained results grouped by Class (C), 5-point Likert scale levels (from 1 to 5), and the number of respondents. We report the average and the mode of the perceived severity. 
Most of the respondents who perceived a problem in the inspected classes considered it at least with a \textit{medium severity}. 

Table~\ref{tab:SpearmanSeverity} presents Spearman's $\rho$ to test H2. We note that correlation is statistically significant for both CyC and CoC. Therefore, we can affirm that \textbf{the complexity metrics showed a poor but statistically significant negative correlation.}




\begin{figure}
    \centering
\includegraphics[width=\linewidth]{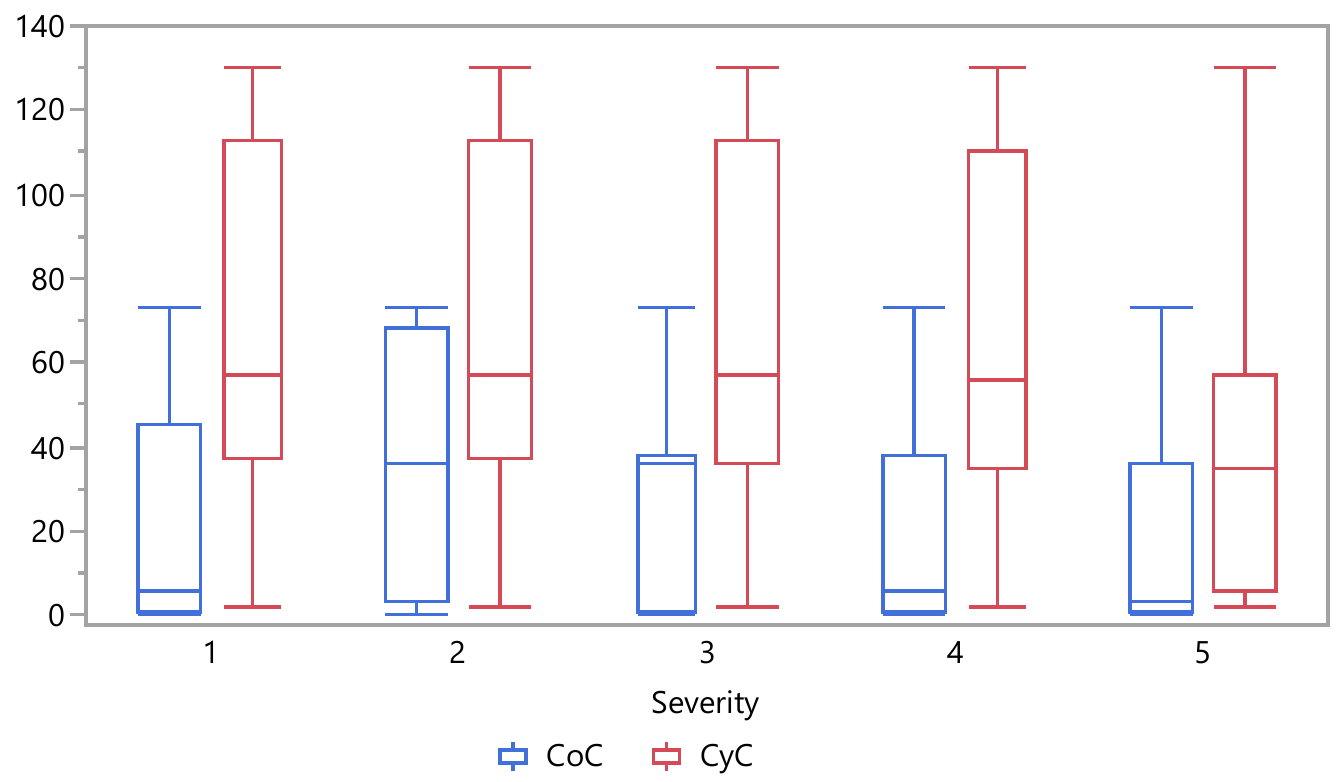}
    \caption{Distribution of complexity metrics values by the perceived severity of the design problem - (RQ$_3$)} 
    \label{fig:BoxPlotSeverity}
\end{figure}

%% file: Tables/understendability.tex
\begin{table}
\centering
\footnotesize
\caption{Perceived Code Understandability (RQ$_1$)}
\label{tab:ResultsRQ1}
\begin{tabular}{c|l|r|p{0.6cm}}
\hline
\textbf{Class} & \textbf{Perc. Understandability} & \textbf{\# respondent}& \textbf{Mode} \\ \hline
\multirow{5}{*}{C1} &  1-Very Hard&  20 & \multirow{5}{*}{3} \\
& 2-Hard & 54  &\\
& 3-Neither Hard or Easy & 76 &\\
& 4-Easy  &  42 &\\ 
& 5-Very Easy &  8  & \\ 

\multirow{5}{*}{C2} &  1-Very Hard& 48   & \multirow{5}{*}{2}  \\
& 2-Hard  &  55 &\\
& 3-Neither Hard or Easy &  58 &\\
& 4-Easy  &  28 &\\ 
& 5-Very Easy &  3 & \\ 
   
\multirow{5}{*}{C3} & 1-Very Hard&  51 &  \multirow{5}{*}{2}\\
& 2-Hard & 49 &\\
& 3-Neither Hard or Easy &  58 &\\
& 4-High  & 29 &\\ 
& 5-Very Easy &  7 & \\ 
   
\multirow{5}{*}{C4} &   1-Very Hard&  27 &  \multirow{5}{*}{3} \\
& 2-Hard &  46 &\\
& 3-Neither Hard or Easy &  65 &\\
& 4-Easy  &  40 &\\ 
& 5-Very Easy &  12 & \\ 
   
\multirow{5}{*}{C5} &   1-Very Hard&  22 & \multirow{5}{*}{4} \\
& 2-Hard &  35 &\\
& 3-Neither Hard or Easy &  67 &\\
& 4-Easy  &  49 &\\ 
& 5-Very Easy &  23 & \\ 
   
\multirow{5}{*}{C6} & 1-Very Hard& 8  & \multirow{5}{*}{3}  \\
& 2-Hard &  21 &\\
& 3-Neither Hard or Easy &  34 &\\
& 4-Easy  & 72 &\\ 
& 5-Very Easy &  60 & \\ 

\multirow{5}{*}{C7} & 1-Very Hard& 18 & \multirow{5}{*}{4}\\
& 2-Hard &  35 &\\
& 3-Neither Hard or Easy &  61 &\\
& 4-Easy  &  52 &\\ 
& 5-Very Easy &  29 & \\ 
   
\multirow{5}{*}{C8} &  1-Very Hard&  7 & \multirow{5}{*}{4} \\
& 2-Hard &  14 &\\
& 3-Neither Hard or Easy &  39 &\\
& 4-Easy  &  50 &\\ 
& 5-Very Easy &  83 & \\ 
   
\multirow{5}{*}{C9} &  1-Very Hard& 29  & \multirow{5}{*}{4} \\
& 2-Hard &  29 &\\
& 3-Neither Hard or Easy &  38 &\\
& 4-Easy  &  35 &\\ 
& 5-Very Easy &  66 & \\ 
   
\multirow{5}{*}{C10} &  1-Very Hard&  10   & \multirow{5}{*}{4}  \\
& 2-Hard &  18 &\\
& 3-Neither Hard or Easy &  35 &\\
& 4-Easy  &  52 &\\ 
& 5-Very Easy&  75 & \\ 

\multirow{5}{*}{C11} & 1-Very Hard& 6 & \multirow{5}{*}{4}\\
& 2-Hard &  10 &\\
& 3-Neither Hard or Easy &  31 &\\
& 4-Easy  &  57 &\\ 
& 5-Very Easy &  88 & \\ 
   
\multirow{5}{*}{C12} & 1-Very Hard&  6 &  \multirow{5}{*}{4} \\
& 2-Hard &  19 &\\
& 3-Neither Hard or Easy &  28 &\\
& 4-Easy  & 51 &\\ 
& 5-Very Easy &  92 & \\\hline 
\end{tabular}
\end{table}

%% file: Tables/severity.tex
\begin{table}
\centering
\footnotesize
\caption{Problem description and Perceived Severity (RQ$_2$)}
\label{tab:ResultsRQ2RQ3}
\begin{tabular}{p{0.5cm}|p{0.4cm}|p{0.4cm}|p{0.4cm}|p{0.4cm}|p{2.7cm}|p{0.3cm}|p{0.3cm}}
\hline 


\multirow{2}{*}{\textbf{Class}} &  \multicolumn{2}{c|}{\textbf{Perceived}}	&\multicolumn{2}{c|}{\textbf{Described}} &\multicolumn{2}{c|}{\textbf{Severity}} & \multirow{2}{*}{\textbf{Mode}} \\ \cline{2-7}

& \textbf{\#} & \textbf{\%} & \textbf{\#}& \textbf{\%}  & &{\textbf{\#}} &\\ \hline 

\multirow{5}{*}{C1} &	\multirow{5}{*}{146}	&	\multirow{5}{*}{69}	&	\multirow{5}{*}{146}	&	\multirow{5}{*}{100} &	1-Very High Severity&	5	&	\multirow{5}{*}{3}	\\
&		&				&		&		&	2-High Severity	&	11	&		\\
&		&				&		&		&	3-Neither Low or High&	25	&		\\
&		&			&		&		&	4-Low Severity	&	24	&		\\
&		&				&		&		&	5-Very Low Severity&	16	&		\\  

\multirow{5}{*}{C2}	&	\multirow{5}{*}{159}	&	\multirow{5}{*}{75}	&	\multirow{5}{*}{159}	&	\multirow{5}{*}{100}	&	1-Very High Severity&	7	&	\multirow{5}{*}{3}	\\
	&	&		&				&		&	2-High Severity	&	29	&		\\
	&	&		&			&		&	3-Neither Low or High&	23	&		\\
    &		&				&		&		&	4-Low Severity	&	21	&		\\
	&	&				&		&		&	5-Very Low Severity	&	8	&		\\  

\multirow{5}{*}{C3}	& 	\multirow{5}{*}{154}	&	\multirow{5}{*}{74}	&	\multirow{5}{*}{154}	&	\multirow{5}{*}{100}		&	1-Very High Severity&	9	&	\multirow{5}{*}{3}	\\
&		&		&				&		&	2-High Severity	&	21	&		\\
&		&		&			&		&	3-Neither Low or High&	35	&		\\
&		&		&				&		&	4-Low Severity	&	16	&		\\
&		&		&			&		&	5-Very Low Severity	&	10	&		\\ 

\multirow{5}{*}{C4}	& 	\multirow{5}{*}{127}	&	\multirow{5}{*}{60}	&	\multirow{5}{*}{127}	&	\multirow{5}{*}{100}		&	1-Very High Severity	&	8	&	\multirow{5}{*}{3}	\\
&		&		&				&		&	2-High Severity	&	17	&		\\
	&	&		&			&		&	3-Neither Low or High&	17	&		\\
&		&		&			&		&	4-Low Severity	&	21	&		\\
&		&		&				&		&	5-Very Low Severity	&	12	&		\\

\multirow{5}{*}{C5}	& 	\multirow{5}{*}{145}	&	\multirow{5}{*}{69}	&	\multirow{5}{*}{145}	&	\multirow{5}{*}{100}	&	1-Very High Severity&	11	&	\multirow{5}{*}{3}	\\
&		&		&				&		&	2-High Severity	&	17	&		\\
	&	&		&			&		&	3-Neither Low or High&	21	&		\\
 &    &		&			&		&	4-Low Severity	&	22	&		\\
	&	&		&			&		&	5-Very Low Severity	&	15	&		\\

\multirow{5}{*}{C6}	 &	\multirow{5}{*}{81}	&	\multirow{5}{*}{38}	&	\multirow{5}{*}{81}	&\multirow{5}{*}{100}		&	1-Very High Severity&	4	&	\multirow{5}{*}{4}	\\
&		&		&				&		&	2-High Severity	&	5	&		\\
&		&		&				&		&	3-Neither Low or High&	9	&		\\
&		&		&				&		&	4-Low Severity	&	11	&		\\
	&	&		&				&		&	5-Very Low Severity	&	21	&		\\ 

\multirow{5}{*}{C7}	&	\multirow{5}{*}{149}	&	\multirow{5}{*}{71}	&	\multirow{5}{*}{116}	&	\multirow{5}{*}{78}		&	1-Very High Severity&	9	&	\multirow{5}{*}{3}	\\
	&	&		&				&		&	2-High Severity	&10	&		\\
&		&		&			&		&	3-Neither Low or High&	24	&		\\
	&	&		&				&		&	4-Low Severity	&	24	&		\\
	&	&		&				&		&	5-Very Low Severity	&	13	&		\\

\multirow{5}{*}{C8}	 &	\multirow{5}{*}{60}	&	\multirow{5}{*}{28}	&	\multirow{5}{*}{60}	&	\multirow{5}{*}{100}		&	1-Very High Severity&	0	&	\multirow{5}{*}{4}	\\
	&	&		&				&		&	2-High Severity	&	4	&		\\
	&	&		&			&		&	3-Neither Low or High&	8	&		\\
	&	&		&				&		&	4-Low Severity	&	8	&		\\
	&	&		&				&		&	5-Very Low Severity	&	17	&		\\  

\multirow{5}{*}{C9}	 &	\multirow{5}{*}{96}	&	\multirow{5}{*}{46}	&	\multirow{5}{*}{75}	&	\multirow{5}{*}{78}		&	1-Very High Severity	&	6	&	\multirow{5}{*}{4}	\\
&		&		&				&		&	2-High Severity	&	9	&		\\
&		&		&				&		&	3-Neither Low or High&	11	&		\\
&		&		&			&		&	4-Low Severity	&	9	&		\\
&		&		&			&		&	5-Very Low Severity&	20	&		\\ 

\multirow{5}{*}{C10}	&	\multirow{5}{*}{39}	&	\multirow{5}{*}{19}	&	\multirow{5}{*}{39}	&	\multirow{5}{*}{100}	&	1-Very High Severity	&	1	&	\multirow{5}{*}{4}	\\
&		&		&				&		&	2-High Severity	&	4	&		\\
&		&		&				&		&	3-Neither Low or High&	5	&		\\
&		&		&				&		&	4-Low Severity	&	3	&		\\
&		&		&				&		&	5-Very Low Severity	&	10	&		\\ 

\multirow{5}{*}{C11}	&	\multirow{5}{*}{53}	&	\multirow{5}{*}{25}	&	\multirow{5}{*}{53}	&	\multirow{5}{*}{100}		&	1-Very High Severity	&	1	&	\multirow{5}{*}{5}	\\
	&	&		&				&		&	2-High Severity	&	0	&		\\
&		&		&				&		&	3-Neither Low or High&	5	&		\\
	&	&		&				&		&	4-Low Severity	&	9	&		\\
	&	&		&				&		&	5-Very Low Severity	&	16	&		\\ 

\multirow{5}{*}{C12}	&	\multirow{5}{*}{89}	&	\multirow{5}{*}{42}	&	\multirow{5}{*}{55} 	&	\multirow{5}{*}{62}		&	1-Very High Severity&	1	&	\multirow{5}{*}{4}	\\
&		&		&				&		&	2-High Severity	&	6	&		\\
&		&		&				&		&	3-Neither Low or High&	10	&		\\
&		&		&		&		&	4-Low Severity	&	15	&		\\
&		&		&			&		&	5-Very Low Severity	&	20	&		\\\hline 
\end{tabular}
\end{table}

%% file: Section/Discussion.tex
\section{Discussion}
\label{Discussion}
This section presents the discussions and key takeaways from the results of our study.


Regarding \textbf{RQ$_1$}, the accuracy metrics and the IRA deliver an interesting take on early-career researcher proficiency in detecting problematic classes. For instance, high Precision and low Recall mean developers fail to identify problematic classes and overlook them. Identifying non-obvious, more subtle issues likely requires more practice and experience. Hence, industries should strengthen their internal training programs to help early-career developers reach a common ground needed by the specific industry. For instance, since the level of agreement among the developers is quite variable in the cases presented, closer mentorship and regular supervision are needed to develop a more homogenized understanding of problematic classes.

Therefore, team composition should mix early- and mid-career developers when assigning code-reviewing tasks to balance their strengths and weaknesses. Finally, Workshops and hands-on coding sessions for life-long learning introduce some balance into performance metrics for precision and recall over time.

\begin{keyTakeAways}[The role of early-career developers]
Early-career developers should not be the only ones responsible for code-reviewing tasks due to their scarce experience.
\end{keyTakeAways}

Considering RQ$_2$, it was evident that less complex classes were deemed easy to understand, indicating that low CyC and CoC support the understandability of the code. However, if CyC or CoC was high, the opinion on understandability was varied. This is a very interesting result and requires further investigation. 
What is especially eye-opening is that perceptions of understandability varied, with both complexities rated highly. Understandability seems to be somewhat more correlated with CoC. However, the difference in CyC was not drastic. Developers showed greater consensus on CoC as a complexity measure, indicating it might be more useful than the two.
Prior results~\cite{Lavazza2022, Scalabrino2019, Feigenspan2011} indicated that the metrics themselves do not indicate understandability and that the different proposed metrics are not positively correlated~\cite{Lavazza2022}.

\begin{keyTakeAways}[CyC an CoC as understandability indicator]
    Based on our findings, low complexity measures indicate good understandability, but having either CoC or CyC high makes understandability unpredictable.
\end{keyTakeAways}


Finally, regarding RQ$_3$ we found that increased complexity heightened the perception of severity. However, there was significant variance. Early-career developers had a good understanding of the code issues. When a class was identified as having a design problem, developers could also describe the problem. This indicates that highlighting the issues contributing to complexity measures can help maintain high understandability. However, understandability does not depend on the type of problem. Therefore, developers should take minor design issues more seriously.

\begin{keyTakeAways}[CyC an CoC as severity indicator]
    Based on our findings, there is no evidence that CyC or CoC are indicators of early-career perceived severity.
\end{keyTakeAways}

%% file: Section/ThreatsToValidity.tex
\section{Threats to Validity}
\label{Threats}
In this Section, we introduce the threats to validity and discuss the different tactics adopted to mitigate them. We followed the structure suggested by Yin~\cite{Yin2014}, reporting construct validity, internal validity, external validity, and reliability. Moreover, we will also discuss the different tactics adopted to mitigate them.


\textbf{Construct validity.}
Concerning the set of tasks, we considered classes whose code complexity was measured by the same tool that allows us to compute both complexities considered in this work.
We checked each question to avoid potential misunderstandings, negative questions, and threats. The perceived priority of the design problem was collected by asking the participants to describe the problem they perceived to understand if their perception is related to the identified problem and not to other potential issues in the code.
We asked the participants to rate the severity of the problem using a Likert scale to allow us to compare the responses based on a homogeneous scale. To reduce this threat, we checked the correctness of the identification both manually and using automated tools.   

\textbf{Internal Validity.}
Considering the respondents, we selected junior developers with a maximum of 4 years of experience in programming skills to better focus on our goal.
However, we are aware that the results could be biased by selecting participants from a set of developers more deeply trained in these tasks. 
We adopted a treatment set of classes to measure whether using the same tool to measure CoC and CyC complexities was possible. 
We know that further studies with different analyzed classes, considering the missing groups, are needed to confirm our results.

Moreover, one class from one domain (e.g., accounting) may be very understandable to someone with experience developing software for that domain but not at all comprehensible to someone without experience. So, the two developers' answers would differ for the same class. 
Considering the amount of code used in the study (12 units of code), acknowledging that it might be too small if the studied phenomenon is large (e.g., the architecture of a large software product). In our case, we reasoned that the studied phenomenon manifests within small pieces of code. We selected 12 classes that contain the studied phenomena (see Table~\ref{tab:Classes}), which are examined by 216 developers. The fact that the same phenomena were studied 216 times 12 gives us confidence that---in our context---the data set is large enough to yield representative results.

\textbf{External validity.} We considered only junior developers with a maximum of four years of experience. We know developers with more experience might have a different perception of code complexity and understandability due to their familiarity with more complex code. Therefore, our results can not be generalized across the entire population of developers.

\textbf{Reliability.}
Three experts checked the survey on empirical studies. Moreover, it was ensured that the subjects of both groups had similar backgrounds and knowledge regarding code understandability and inspection.  

%% file: Section/Conclusion.tex
\section{Conclusion}
\label{Conclusion}
We designed and conducted a case study among 216 early-career developers. We asked them to manually inspect 12 Java classes that exhibit different code complexity levels as Cognitive and CyC measured by SonarQube. For each class, developers had to rate the code's understandability.
CoC and CyC are medium negatively correlated with code understandability and poor negatively correlated with the severity of code problems. We expected to find more problems in classes with higher levels of complexity, mainly because we were expecting these classes to be harder to understand. Therefore, we cannot claim that classes with a higher CyC or CoC are affected by more severe problems than those with lower levels of complexity. Moreover, we also suggest, based on our results, that early-career developers should not be the only ones responsible for code-reviewing tasks due to their scarce experience.
Future work will involve replicating this study with more developers to suggest refactoring actions for identified problems. We will also compare the perceived understandability of code between junior and senior developers and explore other programming languages like Python and JavaScript. Finally, we plan to use larger, more varied code samples, control for variables such as domain knowledge and employ different tools to measure complexity.